\documentclass[preprint, superscriptaddress,preprintnumbers,amsmath,amssymb,pra]{revtex4}
\usepackage{graphicx}

\begin{document}

\thispagestyle{empty}

\title{Optical chopper driven by the Casimir force}

\author{
G.~L.~Klimchitskaya}
\affiliation{Central Astronomical Observatory at Pulkovo of the
Russian Academy of Sciences, Saint Petersburg,
196140, Russia}
\affiliation{Institute of Physics, Nanotechnology and
Telecommunications, Peter the Great Saint Petersburg
Polytechnic University, Saint Petersburg, 195251, Russia}

\author{
V.~M.~Mostepanenko}
\affiliation{Central Astronomical Observatory at Pulkovo of the
Russian Academy of Sciences, Saint Petersburg,
196140, Russia}
\affiliation{Institute of Physics, Nanotechnology and
Telecommunications, Peter the Great Saint Petersburg
Polytechnic University, Saint Petersburg, 195251, Russia}
\affiliation{Kazan Federal University, Kazan, 420008, Russia}

\author{
V.~M.~Petrov}
\affiliation{Institute of Advanced Manufacturing Technologies,
Peter the Great Saint Petersburg
Polytechnic University, Saint Petersburg, 195251, Russia}

\author{
T.\ Tschudi}
\affiliation{Institute of Applied Physics, Darmstadt  University of
Technology, Hochschulstrasse 6, Darmstadt 64289, Germany}

\begin{abstract}
We propose the experimental scheme and present detailed theory
of the optical chopper which functionality is based on the
balance between the Casimir and light pressures. The proposed
device consists of two atomically thin metallic mirrors forming
the Fabry-P\'{e}rot microfilter. One of the mirrors is deposited
on a solid cube and another one on a thinner wall subjected to
bending under the influence of the attractive Casimir force and
repulsive force due to the pressure of light from a continuous
laser amplified in the resonator of a microfilter. The separation
distance between the mirrors should only slightly exceed the
half wavelength of the laser light. It is shown that in this
case the resonance condition in the microfilter alternatively
obeys and breaks down resulting in the periodic pulses of the
transmitted light. The Casimir pressure is calculated taking
into account an anisotropy of the dielectric permittivity of
a metal at several first Matsubara frequencies. The reflectivity
properties of atomically thin metallic mirrors in the optical
spectral range are found using the experimentally consistent
phenomenological approach developed earlier in the literature.
The specific values of all parameters, found for the
microfilter made of quartz glass with Ag mirrors, demonstrate
its workability. The proposed optical chopper may find
prospective applications in the emerging field of nanotechnology
exploiting the effects of quantum fluctuations.
\end{abstract}

\maketitle

\section{INTRODUCTION}

In the last few decades, advances in the integrated-circuit
fabrication techniques allowed producing of microelectomechanical
(MEMS) and nanoelectromechanical (NEMS) systems with sizes ranging
from micrometers to nanometers \cite{1}. It is common knowledge
that MEMS and NEMS devices are increasingly used in optical and
cellular communications, as well as in a variety of sensors and in
many other applications. As was noticed more than thirty years ago
\cite{2,3}, with shrinking MEMS dimensions to submicrometer level,
in addition to electric forces, the van der Waals \cite{4} and
Casimir \cite{5} forces induced by the electromagnetic fluctuations
come into play. These forces act between uncharged material surfaces
and become dominant at separations of several nanometers and several
hundred nanometers, respectively. In an early stage, the combined
action of the Casimir and elastic forces in MEMS devices has been
studied in Ref. \cite{6}. Later on the role of roughness and
electrostatic effects was also considered in Refs.~\cite{7,8}.

Experimentally the combined role of the electrostatic and Casimir
forces in the loss of functionality of MEMS device, when the moving
part of it jumps to a fixed electrode, was investigated in
Refs.~\cite{9,10}. This phenomenon was called a {\it pull-in} or
{\it stiction}.
A short time later it was experimentally demonstrated that the
Casimir force is not only detrimental to MEMS and NEMS functionality,
but can be also used for actuation of microdevices in place of the
electric force \cite{11,12}. The original device created in
Refs.~\cite{11,12} has been called a
{\it micromechanical Casimir oscillator}.
In the next few years this device was refined and actively exploited
for both precise measurements of the Casimir interaction in
fundamental physics and for creation of novel MEMS and NEMS
(see Refs.~\cite{13,14} for a review).

In recent years a lot of high-precision experiments on measuring
the Casimir interaction between smooth surfaces of metallic
\cite{15,16,17,18,19,20,21,22,23} and semiconductor
\cite{24,25,26,27,28,29,30,31,32,33,34}
test bodies have been performed by means
of an atomic force microscope and a micromechanical oscillator.
In several Casimir experiments the structured (sinusoidally and
rectangular corrugated) test bodies have also been used
\cite{35,36,37,38,39}. All this gave impetus to diverse applications of the
obtained results to MEMS and NEMS devices driven by the Casimir
force. Thus, the role of geometry and dielectric properties of
materials in the stability of Casimir-actuated nanodevices was
analyzed in Refs.~\cite{40,41}. The actuation of MEMS under the
influence of Casimir force with account of surface roughness and
amorphous to crystalline phase transformations was investigated
\cite{42,43,44}. The Casimir forces on a silicon micromechanical
chip have been experimentally demonstrated in Refs.~\cite{45,46}.
In Ref.~\cite{47} the method was suggested on how to control the
mechanical switch using an enhancement of the Casimir force
between a graphene sheet and a silicon membrane. We note, however,
that the scheme of the Casimir switch suggested in Ref.~\cite{48}
as a possibility to significantly alter the optical output rate
by the vacuum force is inoperative. As the authors themselves
recognize, measurements of the Casimir force in the separation
region from 0.7 to 2\,nm, required in their scheme, are
challenging. Of even greater concern is the fact that
Ref.~\cite{48} uses an ideal-metal expression for the Casimir force
between gold-coated surfaces at so short separations and, thus,
overestimates the force magnitude by at least a factor of twenty
\cite{49}.

In this paper, we propose the possibility to create the
optical chopper driven by the Casimir force.
The feasibility of our proposal is supported
with detailed theory. The key element of the proposed  setup is the
SiO${}_2$ microdevice incorporating the Fabry-P\'{e}rot microfilter
with two parallel thin metallic mirrors  which form
a microresonator. The left mirror should be deposited on the side of
a solid SiO${}_2$ cube and the right one on a relatively
thin SiO${}_2$ wall
subjected to bending under the influence of the Casimir force
acting between the two mirrors in high vacuum. Note that
detection of the mechanical deformation of a macroscopic object
induced by the Casimir force was made in Refs.~\cite{50,51}
by means of an adaptive holographic interferometer. The
length of a resonator cavity (i.e., the separation distance
between the foot parts of the mirrors) should be made only slightly
larger than the half wavelength $\lambda/2$ of the laser light
incident from the left.
In the absence of laser light, the Casimir force shifts the top of
the right mirror  slightly closer to the top of the
left one than their foot parts. This is a stable position where the
Casimir force is balanced by the restoring elastic force.
As a result, when the laser is switched on and the light beam
enters a microfilter through the SiO${}_2$ cube and the left
mirror, the effective resonator length at the beam section
will be approximately
equal to $\lambda$/2. This leads to a cyclic process. First,
the amplitude of a standing wave in the filter resonator will
instantaneously increase resulting in detection of a relatively
high level of intensity of the transmitted light.
The repulsive force due to the
light pressure in the resonator will  compensate the
Casimir force and the right mirror will become vertical.
This is an unstable position where the elastic force vanishes.
The effective resonator length here is larger than
$\lambda$/2 violating the resonance condition. Then the wave
amplitude in the gap will fall down leading to
almost zero level of intensity of the transmitted light.
Finally, the Casimir force, which will be not balanced by the
repulsive force due to
light pressure any more, will return the right mirror to its
initial position where the wall is slightly tilted to the left.
Here, the resonance condition for the
incident light beam is again obeyed with sufficient precision, and
the next cycle starts.

We emphasize that the proposed microresonator should not be
considered as an optomechanical cavity which usually consists of a
fixed mirror and mechanical oscillator, i.e., another mirror is
attached to a spring (see the monograph \cite{51aa} and review
\cite{51bb}). The point is that for optomechnanical cavities the
dynamics of the second mirror is important for the functionality of
a device. This is, however, not the case for the proposed optical
chopper which is driven not by a mechanical force or a mechanical
force in combination with the light pressure, but mostly by the
Casimir force acting on the right mirror deposited on a wall. This
is reached due to the chosen initial position of the foot part of the
 right mirror at more than $\lambda/2$ separation
from the left one, i.e., at a
distance where the resonance condition is violated.
In this case one does not need to know a detailed dynamics of
the system to prove an existence of the cyclic process
(see Secs.~II and IV for more details including the role of
optomechanical effects).

We have developed theoretical description of the physical
processes in the described above experimental setup. The Casimir force
is calculated on the basis of the Lifshitz theory at nonzero
temperature with account of an anisotropy of the dielectric
permittivities for thin metallic films, forming the resonator
mirrors, at several first Matsubara frequencies \cite{51a}.
The reflectance and transmittance of mirrors forming the
Fabri-P\'{e}rot filter at the used laser wavelength are found with due
regard to increased transparency of atomically thin metallic films \cite{51b}.
The light pressure in the resonator is also calculated.
 The balance between
calculated Casimir and light pressures enables one to predict the
formation of pulses in the transmitted light.
We argue that the designed device is advantageous as compared to
the commonly used mechanical optical choppers exploiting the
 wheels of various shape which should have a highly
stable rotating speed.

The paper is organized as follows. In Sec.~II, we present
some details of the proposed setup.
Section III contains  calculations of the Casimir
pressure in the experimental configuration.
In Sec.~IV we find the force due to the
light pressure and its balance with the Casimir force.
In Sec.~V the reader will find our conclusions
and a discussion.

\section{The proposed setup}

\begin{figure}[b]
\vspace*{1cm}
\centerline{\hspace*{2.5cm}
\includegraphics{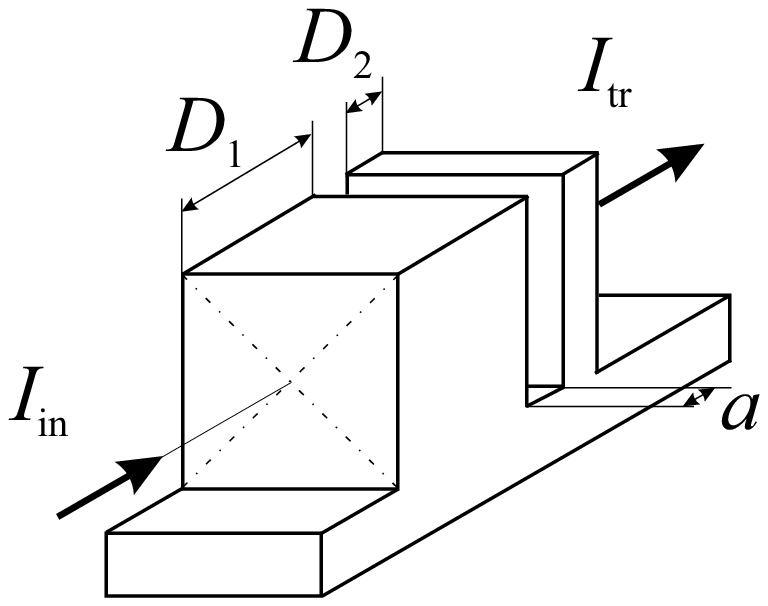}
}
\vspace*{-19cm}
\caption{\label{fg1}
Schematic of the SiO${}_2$ microdevice incorporating the microfilter
of Fabry-P\'{e}rot with length $a$ (see text for further
discussion).}
\end{figure}

The heart of the proposed optical chopper driven by the Casimir force is
 a SiO${}_2$  microdevice
which can be manufactured using the  technique of ion-beam etching.
The main part of this device is  the Fabry-P\'{e}rot microfilter
formed by a cube with the side $D_1=50\,\mu$m and  a square
wall of thickness $D_2=5\,\mu$m of the same side-length.
Both the cube and the wall are located at the joint
base parallel to each other (see Fig.~\ref{fg1}).
The right face of the cube and the left face of the wall in Fig.~\ref{fg1}
should be coated by thin metallic layers
to form the resonator mirrors of the filter.
The thickness of the mirrors $d$ could be in the range from 0.5 to 3\,nm.
The resonator length (i.e., the distance between metallic mirrors) is
notated $a$.

Some details on how to fabricate microdevices like that
one shown in Fig.~1 can be found in the literature
on measuring the Casimir force and its applications
in nanotechnology. Although fabrication of large area
microstructures with a uniform gap of several hundred nanometers
width and vertical sidewalls remains challenging \cite{39}, the
nanofabrication processes were developed allowing to produce two
interacting surfaces that are automatically aligned and almost
parallel to each other \cite{39,45}. In so doing, with the etch mask
defined by electron-beam lithography, a high degree of parallelism
is ensured \cite{45}. The deposition of metallic mirrors can be
performed by either electroplating or sputtering \cite{39}. Another
technology of producing a microdevice shown in Fig.~1 suggests
manufacturing the two halves of this device separately. Then both
halves should be placed into a vacuum chamber where the metallic
mirrors are deposited. The assembly and alignment of the entire
device should be made inside the vacuum chamber using the high
precision positioning technology described in Refs.~\cite{50,51,53a}.

As a source of light
of intensity $I_{\rm in}$ incident on the SiO${}_2$ microdevice
in the proposed experiment,  a CW Nd-YAG laser with a wavelength
of the second harmonic $\lambda=532.0\,$nm can be used.
It is suggested to fabricate the resonator with length equal to
$a=\lambda/2+\Delta\lambda$ where $\Delta\lambda$ is sufficiently large
to break down the resonance condition
$a\approx\lambda/2$ in the absence of any external force.
For smaller $a$, i.e., for $|a-\lambda/2|<\Delta\lambda$,
the resonance condition is assumed to be obeyed with sufficient
precision (see Sec.~IV for the specific values of $\Delta\lambda$).

\begin{figure}[b]
\vspace*{-0cm}
\centerline{\hspace*{2.5cm}
\includegraphics{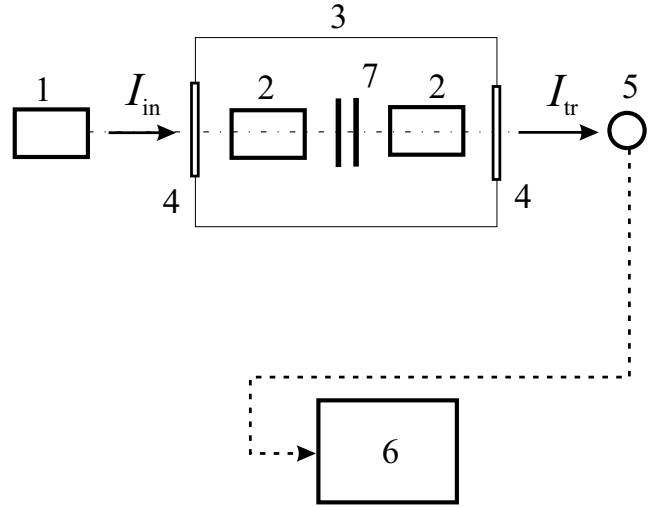}
}
\vspace*{-17cm}
\caption{\label{fg2}
General scheme of the optical chopper driven by the Casimir
force: 1 --- laser, 2 --- beam-forming systems, 3 --- vacuum
chamber, 4 --- optical windows, 5 --- photodetector, 6 ---
two-channel oscilloscope, 7 --- Fabry-P\'{e}rot microfilter.}
\end{figure}

The SiO${}_2$ system should incorporate two pairs of the optical
windows and  beam-forming
systems inserted into the vacuum chamber (see Fig.~\ref{fg2})
at sufficiently low pressure of about $10^{-6}-10^{-7}\,$Torr.
The beam-forming systems are needed to form the light beams having
the Gaussian-like profiles
with the diameter equal to approximately $40\,\mu$m for the wavelength
of 532.0\,nm.
The transmitted light beam can be detected by a
photodetector (see Fig.~\ref{fg2}). Special attention should be paid to the
stability of the setup. For this purpose, it is desirable to
deposit the vacuum chamber on the optical table with
an active air-pumped stabilization.

It is expected that the photodetector will register
pulses  of light transmitted through
the Fabri-P\'{e}rot microfilter, i.e., the high level of intensity
of the transmitted light will alternate with periods when the intensity
of transmitted light reduces to almost zero.
According to Sec.~I, this expectation is based on
 the action of the Casimir force. When the laser is switched off,
this force slightly tilts the wall
of our microdevice in the direction of the left mirror. As  a result, the
separation distance between the mirrors decreases and
the resonance condition $a\approx\lambda/2$
is obeyed with  sufficient precision. In this initial position the Casimir
force is balanced by the restoring elastic force. After the laser is
switched on, an amplitude of the standing wave in the gap between
the mirrors will instantaneously
increase taking into account rather high quality factor of the resonator
(see Sec.~IV). Because of this, the photodetector will detect high level of
the transmitted light, and the repulsive force due to the
light pressure in the gap will compensate the attractive Casimir
force. Thus, the wall will take the vertical position where it is separated
from the left mirror by the distance $a=\lambda/2+\Delta\lambda$.
In this position the elastic force is equal to zero but the resonance
condition breaks down leading to an instantaneous drop
 of the wave amplitude in the gap  and to
low level of intensity of the transmitted light.
At this stage, the attractive Casimir force returns the wall to its initial
(tilted) position, where it is balanced by the elastic force. In this
position the resonance condition $a\approx\lambda/2$ is
again obeyed with sufficient precision, and the next cycle starts.

In Secs.~III and IV these qualitative  considerations receive
quantitative confirmation by using
the Lifshitz theory of the Casimir force and calculating the parameters
of a microfilter.
As already noted in Sec.~I, the proposed microresonator
should not be considered as an optomechanical cavity
which incorporates a mechanical oscillator and gives rise
to the process of light-induced cyclic response, i.e.,
self-excited oscillation. This is already seen from the fact that,
in the absence of the Casimir force, no oscillation arises in our
microresonator with the laser light on (because in this case
the right mirror would be vertical and the resonance condition
is, thus, violated).
In the proposed setup, it is only the Casimir force which leads to a
fulfilment of the resonance condition in the initial position
of a cycle by tilting the wall to the left in the absence of laser light,
and it is the light pressure which leads to a violation
of this condition when the wall becomes vertical. Thus, if
it is possible to ensure the balance between the Casimir and light
pressures (see Sec. IV),
the suggested configuration clearly undergoes a cyclic process
even if the
details of an intermediate dynamics remain unknown. Note also that
the dynamics of an optomechanical cavity formed by a stationary mirror
made  of dielectric Si and an oscillating Al mirror was considered in
Ref.~\cite{53b} with account of the Casimir force and Coulomb interaction
due to trapped charges, and a partial agreement between experiment
and theory was reached.

\section{Calculation of the Casimir pressure}

We start from calculation of the Casimir pressure in the configuration of a
microdevice described in Sec.~II. It consists of a SiO${}_2$ cube with
the side $D_1=50\,\mu$m, whose right face is coated by the metallic film
of thickness $d$, and parallel to it SiO${}_2$ wall of the same area.
The thickness of this wall is $D_2=5\,\mu$m. The left face of the wall is
also coated by metallic film of thickness $d$. The separation distance between
the cube and the wall is only slightly larger than $\lambda/2=266\,$nm
(see Fig.~\ref{fg3} and compare it with Fig.~\ref{fg1}).
Taking into account that $D_1\gg a$, one can consider the opposite faces of a cube and
a wall as having the infinitely large area. However, the finite thickness of
the cube, of the wall, and of the metallic coatings should be taken into
account in computations of the Casimir pressure. In so doing, we also take proper
account of the anisotropy of atomically thin metallic film of thickness $d$
which are described as uniaxial crystals \cite{51a}.

The microdevice under consideration is assumed to be at temperature $T$ in
thermal equilibrium with the environment.
In this case the Casimir pressure acting on the wall is given by the following
Lifshitz formula for the four-layer system \cite{5,49,52,53}
\begin{eqnarray}
&&
P(a,T)=-\frac{k_BT}{\pi}\sum_{l=0}^{\infty}{\vphantom{\sum}}^{\prime}
\int_{0}^{\infty}k_{\bot}\,dk_{\bot}\,q_l
\nonumber\\
&&~~
\times
\left\{\left[\frac{e^{2aq_l}}{R_{{\rm TM},l}^{(1)}R_{{\rm TM},l}^{(2)}}
-1\right]^{-1}\right.
\nonumber\\
&&~~~~~
+
\left.\left[\frac{e^{2aq_l}}{R_{{\rm TE},l}^{(1)}R_{{\rm TE},l}^{(2)}}
-1\right]^{-1}\right\}.
\label{eq1}
\end{eqnarray}
\noindent
Here, $k_B$ is the Boltzmann constant, the prime on the summation sign multiplies
the term with $l=0$ by 1/2, $k_{\bot}=|\mbox{\boldmath$k$}_{\bot}|$ is the
magnitude of the projection of the wave vector on the plane of the wall, and
the factor $q_l$ is defined as
\begin{equation}
q_l=\sqrt{k_{\bot}^2+\frac{\xi_l^2}{c^2}},
\label{eq2}
\end{equation}
\noindent
where $\xi_l=2\pi k_BTl/\hbar$ with $l=0,\,1,\,2,\,\ldots$ are the Matsubara
frequencies.

The amplitude reflection coefficients $R_{{\rm TM(TE)},l}^{(i)}$ on the left
($i=1$) and right ($i=2$) plates of our device at the Matsubara frequencies
entering the Lifshitz formula (\ref{eq1}) are defined for two independent
polarizations of the electromagnetic field, transverse magnetic (TM) and
transverse electric (TE) and have the form \cite{5,49,52,53}
\begin{eqnarray}
&&
R_{{\rm TM(TE)},l}^{(i)}\equiv R_{{\rm TM(TE)}}^{(i)}(i\xi_l,k_{\bot})
\nonumber \\
&&
=\frac{r_{{\rm TM(TE)},l}^{(v,m)}+R_{{\rm TM(TE)},l}^{(g,i)}
e^{-2dk_{{\rm TM(TE)},l}^{(m)}}}{1+
r_{{\rm TM(TE)},l}^{(v,m)}R_{{\rm TM(TE)},l}^{(g,i)}
e^{-2dk_{{\rm TM(TE)},l}^{(m)}}}.
\label{eq3}
\end{eqnarray}
\noindent
In this equation, the quantities $k_{{\rm TM(TE)},l}^{(m)}$, related to
anisotropic metallic films, are given by \cite{5,54,55}
\begin{eqnarray}
&&
k_{{\rm TM},l}^{(m)}\equiv k_{{\rm TM}}^{(m)}(i\xi_l,k_{\bot})=
\sqrt{\frac{\varepsilon_{xx,l}^{(m)}}{\varepsilon_{zz,l}^{(m)}}k_{\bot}^2
+{\varepsilon_{xx,l}^{(m)}}\frac{\xi_l^2}{c^2}},
\nonumber\\
&&
k_{{\rm TE},l}^{(m)}\equiv k_{{\rm TE}}^{(m)}(i\xi_l,k_{\bot})=
\sqrt{k_{\bot}^2
+{\varepsilon_{xx,l}^{(m)}}\frac{\xi_l^2}{c^2}},
\label{eq4}
\end{eqnarray}
\noindent
where the components of the diagonal dielectric tensor of a metal are
$\varepsilon_{xx,l}^{(m)}\equiv\varepsilon_{xx}^{(m)}(i\xi_l)=
\varepsilon_{yy}^{(m)}(i\xi_l)$,
$\varepsilon_{zz,l}^{(m)}\equiv\varepsilon_{zz}^{(m)}(i\xi_l)$,
and we assume that the plane $(x,y)$ is parallel to the wall and the $z$
axis is perpendicular to it.

Now we specify the amplitude reflection coefficients entering Eq.~(\ref{eq3}).
The coefficient $r_{{\rm TM(TE)},l}^{(v,m)}$ describes reflection of the
electromagnetic waves at the boundary plane between vacuum and a semispace made
of an anisotropic metal. It takes the form \cite{5,54,55}
\begin{eqnarray}
&&
r_{{\rm TM},l}^{(v,m)}\equiv r_{{\rm TM}}^{(v,m)}(i\xi_l,k_{\bot})=
\frac{\varepsilon_{xx,l}^{(m)}q_l-
k_{{\rm TM},l}^{(m)}}{\varepsilon_{xx,l}^{(m)}q_l+
k_{{\rm TM},l}^{(m)}},
\nonumber \\
&&
r_{{\rm TE},l}^{(v,m)}\equiv r_{{\rm TE}}^{(v,m)}(i\xi_l,k_{\bot})=
\frac{q_l-
k_{{\rm TE},l}^{(m)}}{q_l+
k_{{\rm TE},l}^{(m)}}.
\label{eq5}
\end{eqnarray}
\noindent
The coefficients $R_{{\rm TM(TE)},l}^{(g,i)}$ with $i=1,\,2$ can be presented
similar to Eq.~(\ref{eq3})
\begin{eqnarray}
&&
R_{{\rm TM(TE)},l}^{(g,i)}\equiv R_{{\rm TM(TE)}}^{(g,i)}(i\xi_l,k_{\bot})
\nonumber \\
&&
=\frac{r_{{\rm TM(TE)},l}^{(m,g)}+r_{{\rm TM(TE)},l}^{(g,v)}
e^{-2D_ik_{l}^{(g)}}}{1+
r_{{\rm TM(TE)},l}^{(m,g)}r_{{\rm TM(TE)},l}^{(g,v)}
e^{-2D_ik_{l}^{(g)}}},
\label{eq6}
\end{eqnarray}
\noindent
where
\begin{equation}
k_l^{(g)}\equiv k^{(g)}(i\xi_l,k_{\bot})=
\sqrt{k_{\bot}^2+\varepsilon_l^{(g)}\frac{\xi_l^2}{c^2}},
\label{eq7}
\end{equation}
\noindent
and $\varepsilon_l^{(g)}\equiv\varepsilon^{(g)}(i\xi_l)$ is
the dielectric permittivity of quartz glass SiO${}_2$.

The amplitude reflection coefficients
entering Eq.~(\ref{eq6}) are specified as follows.
The coefficients $r_{{\rm TM(TE)},l}^{(m,g)}$ describe reflection at the
boundary plane between the semispaces made of an anisotropic metal and
a SiO${}_2$ glass. They are given by \cite{5,54,55}
\begin{eqnarray}
&&
r_{{\rm TM},l}^{(m,g)}\equiv r_{{\rm TM}}^{(m,g)}(i\xi_l,k_{\bot})=
\frac{\varepsilon_l^{(g)}k_{{\rm TM},l}^{(m)}-
\varepsilon_{xx,l}^{(m)}k_l^{(g)}}{\varepsilon_l^{(g)}k_{{\rm TM},l}^{(m)}+
\varepsilon_{xx,l}^{(m)}k_l^{(g)}},
\nonumber \\
&&
r_{{\rm TE},l}^{(m,g)}\equiv r_{{\rm TE}}^{(m,g)}(i\xi_l,k_{\bot})=
\frac{k_{{\rm TE},l}^{(m)}-k_l^{(g)}}{k_{{\rm TE},l}^{(m)}+k_l^{(g)}}.
\label{eq8}
\end{eqnarray}

Finally, the coefficients $r_{{\rm TM(TE)},l}^{(g,v)}$ at the boundary plane
between a SiO${}_2$ semispace and vacuum have the form
\begin{eqnarray}
&&
r_{{\rm TM},l}^{(g,v)}\equiv r_{{\rm TM}}^{(g,v)}(i\xi_l,k_{\bot})=
\frac{k_l^{(g)}-\varepsilon_l^{(g)}q_l}{k_l^{(g)}+\varepsilon_l^{(g)}q_l},
\nonumber \\
&&
r_{{\rm TE},l}^{(g,v)}\equiv r_{{\rm TE}}^{(g,v)}(i\xi_l,k_{\bot})=
\frac{k_l^{(g)}-q_l}{k_l^{(g)}+q_l}.
\label{eq9}
\end{eqnarray}

The dielectric permittivity of vitreous SiO${}_2$ at the Matsubara frequencies,
required for computations of the Casimir pressure using Eqs.~(\ref{eq1})--(\ref{eq9}),
was taken from Ref.~\cite{56}.
As the metal of resonator mirrors we have first chosen Au most often used in
precise experiments on measuring the Casimir force
\cite{13,14,15,16,17,18,19,20,21,22,23,24,25,26,27,28,29,30,31,32,33,34,35,36,37,38,39}.
The dielectric tensor of thin Au films consisting of $n=1$, 3, 6, and 15 atomic
layers (i.e., for film thickness of 0.235, 0.705, 1.41, and 3.525\,nm) was found
in Ref.~\cite{51a} within the density functional theory.
It takes into account the effects of anisotropy and uses the optical data of
Ref.~\cite{57} for the complex index of refraction of Au extrapolated to zero
frequency by means of the Drude model. Note that for the atomically thin metallic films
the obtained results do not depend on whether the Drude or the plasma model is
used for extrapolation of the data to zero frequency \cite{54}.

The computational results for the magnitude of negative (attractive) Casimir
pressure were obtained at $T=300\,$K as a function of
the number of atomic layers in the Au films and interpolated in the region of film
thicknesses from 1 to 3.5\,nm.
The calculated Casimir pressures at the separation $a=\lambda/2=266\,$nm are shown by the
bottom line in Fig.~\ref{fg4}. It turns out, however, that as a material of mirrors
in a microfilter Au is rather disadvantageous due to low reflectivity at the
chosen wavelength (see Sec.~IV). Because of this, the computations of the Casimir
pressure were repeated for Ag mirrors. Silver has almost the same lattice parameter
as Au \cite{58}. This allows one to use the same data for
$\varepsilon_{xx}^{(\rm Ag)}/\varepsilon_{\rm isotr}^{(\rm Ag)}$ and
$\varepsilon_{zz}^{(\rm Ag)}/\varepsilon_{\rm isotr}^{(\rm Ag)}$, as were
computed in Ref.~\cite{51a} for Au, and to multiply the dielectric permittivity
of an isotropic (bulk) Ag, $\varepsilon_{\rm isotr}^{(\rm Ag)}$ , found from the
tabulated optical data of Ref.~\cite{57} by this factors.
The computational results  for the magnitude of the Casimir
pressure in the resonator with Ag mirrors
are shown in Fig.~\ref{fg4} as a function of mirror thickness
at $a=\lambda/2=266\,$nm by the top line.

\section{Balance of light and Casimir pressures}

\begin{figure}[b]
\vspace*{-7cm}
\centerline{\hspace*{1.5cm}
\includegraphics{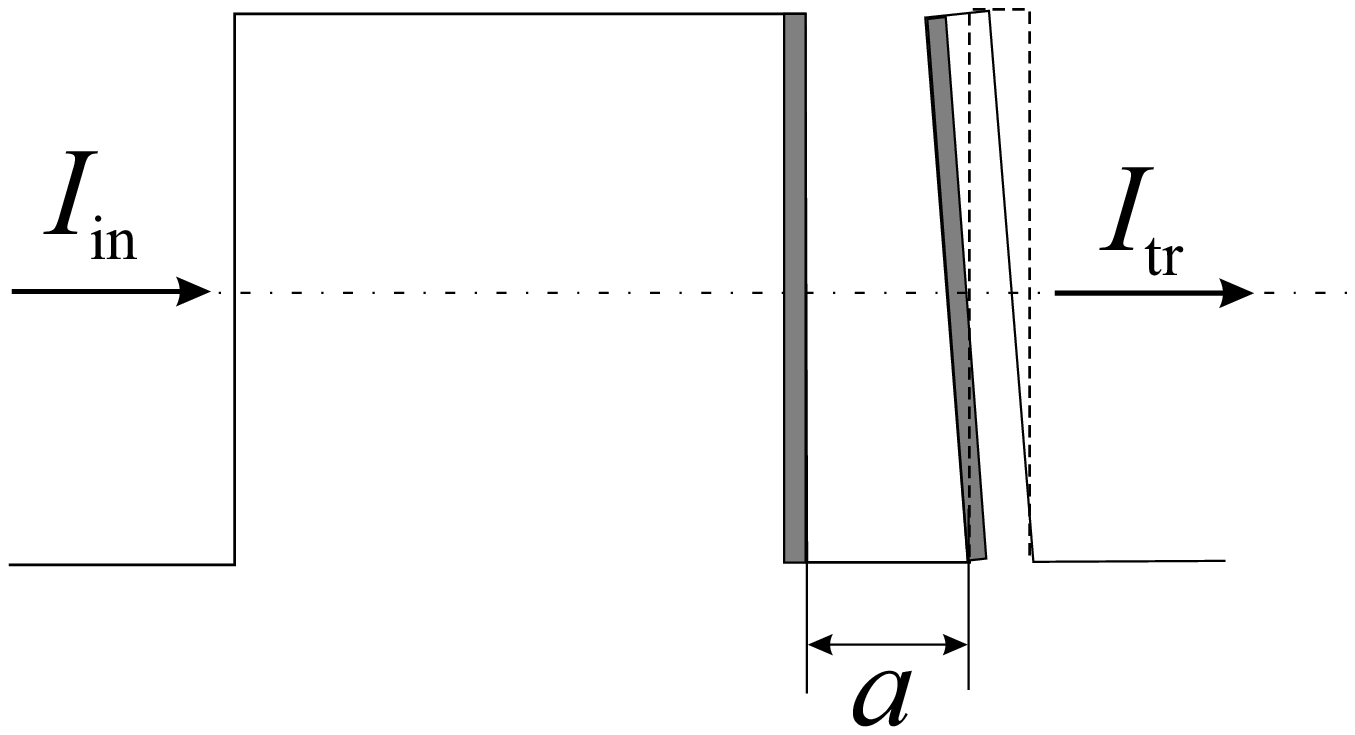}
}
\vspace*{-12cm}
\caption{\label{fg3}
Configuration of the resonator in the Fabry-P\'{e}rot
microfilter. The metallic mirrors are shown not to scale and
marked by dark-grey.}
\end{figure}

\begin{figure}[b]
\vspace*{-7cm}
\centerline{\hspace*{1.5cm}
\includegraphics{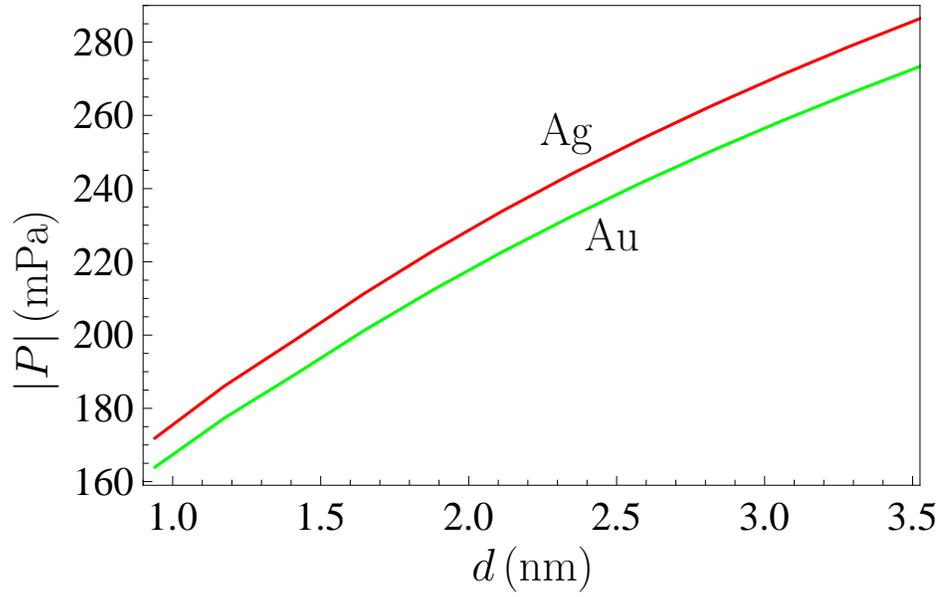}
}
\vspace*{-8cm}
\caption{\label{fg4}
The computational results for the magnitude of the
Casimir pressure in the Fabry-P\'{e}rot microfilter are shown
by the top and bottom lines as functions of mirror thickness
at the separation $a=\lambda/2$ between the mirrors made of
Ag and Au, respectively.}
\end{figure}

The intensity of light incident on the Fabry-P\'{e}rot filter is $I_{\rm in}$
(see Figs.~\ref{fg1} and \ref{fg3}). After the integration of $I_{\rm in}$
over the beam area, one obtains the value of power
$N_{\rm in}\approx 7\,$mW for the chosen laser.
Taking into account high transparency
of quartz glass at the wavelength $\lambda=532\,$nm
($\omega=3.54\times 10^{15}\,$rad/s), one can neglect by the losses in
SiO${}_2$ cube and assume that the same power  $N_{\rm in}$ falls on the
left metallic mirror. According to Fig.~\ref{fg3}, the mirror borders the
glass cube from the left and the vacuum gap from the right. These two media
can be considered as semispaces. Then the magnitude of
the amplitude reflection coefficient
on the metallic film of thickness $d$ at the normal incidence is given by
\cite{59}
\begin{equation}
|R|=\left|\frac{\tilde{r}^{(v,m)}+\tilde{r}^{(m,g)}
e^{-2i\frac{\omega}{c}d\sqrt{\varepsilon^{(m)}}}}{1+
\tilde{r}^{(v,m)}\tilde{r}^{(m,g)}
e^{-2i\frac{\omega}{c}d\sqrt{\varepsilon^{(m)}}}}\right|,
\label{eq10}
\end{equation}
\noindent
where for the complex dielectric permittivity of metal at the frequency
$\omega$ we have
$\sqrt{\varepsilon^{(m)}}=n^{(m)}+i\kappa^{(m)}$.
For the metals used below we have
$n^{(\rm Au)}=0.543$, $\kappa^{(\rm Au)}=2.25$ and
$n^{(\rm Ag)}=0.129$, $\kappa^{(\rm Ag)}=3.19$ \cite{57}.
Note that $\omega$ is sufficiently high, so that the role of anisotropy
discussed in Sec.~III to calculate the Casimir pressure determined by much
lower frequencies is negligibly small. Because of this, the reflection
coefficients $\tilde{r}^{(v,m)}$ and $\tilde{r}^{(m,g)}$ are given by
Eqs.~(\ref{eq5}) and (\ref{eq8}) where one should put
$\varepsilon_{xx}^{(m)}=\varepsilon_{zz}^{(m)}$ and replace $i\xi_l$ with
$\omega$ (we remind that at the normal incidence $k_{\bot}=0$
and the TM and TE
reflection coefficients coincide).

In fact Eq.~(\ref{eq10}) is applicable for sufficiently thick films.
As mentioned in Sec.~I, atomically thin metallic films are characterized
by the increased transparency and this fact should be taken into account in
computations. There are several phenomenological approaches developed to
gain a better understanding of relevant physical mechanisms
(see, e.g., \cite{51b,60,61,62,63,64}). According to the approach of
Refs.~\cite{62,63,64}, for atomically thin metallic films illuminated with
visible light at the normal incidence good agreement with the measurement
results is reached if in the coefficients
$\tilde{r}^{(v,m)}$ and $\tilde{r}^{(m,g)}$
entering Eq.~(\ref{eq10})
one puts $\kappa^{(m)}=0$. In this case we obtain
\begin{equation}
\tilde{r}^{(v,m)}=\frac{n^{(m)}-1} {n^{(m)}+1},
\qquad
\tilde{r}^{(m,g)}=\frac{n^{(g)}-n^{(m)}}{n^{(g)}+n^{(m)}},
\label{eq11}
\end{equation}
\noindent
where $n^{(g)}=\sqrt{\varepsilon^{(g)}}\approx 1.46$ is the real refractive
index of SiO${}_2$ at the used frequency $\omega=3.54\times 10^{15}\,$rad/s.

Substituting Eq.~(\ref{eq11}) in Eq.~(\ref{eq10}), for the reflectance of
metallic film ${\cal R}=|R|^2$ at the frequency $\omega$ one finds
\begin{equation}
{\cal R}=
\frac{\tilde{r}^{(v,m)}{\vphantom{\tilde{r}^{(v)}}}^2+
\tilde{r}^{(m,g)}{\vphantom{\tilde{r}^{(v)}}}^2
e^{-2\alpha d}-
2\tilde{r}^{(v,m)}\tilde{r}^{(m,g)}e^{-\alpha d}\cos\psi}{1+
\tilde{r}^{(v,m)}{\vphantom{\tilde{r}^{(v)}}}^2
\tilde{r}^{(m,g)}{\vphantom{\tilde{r}^{(v)}}}^2
e^{-2\alpha d}{\vphantom{\tilde{r}^{(v)}}}-
2\tilde{r}^{(v,m)}\tilde{r}^{(m,g)}e^{-\alpha d}\cos\psi},
\label{eq12}
\end{equation}
\noindent
where
\begin{equation}
\alpha=\frac{4\pi\kappa^{(m)}}{\lambda}, \qquad
\psi=\frac{4\pi n^{(m)}d}{\lambda}.
\label{eq13}
\end{equation}

In a similar way, for the transmittance of our film we have
\begin{equation}
{\cal T}=
\frac{\tilde{t}^{(v,m)}\tilde{t}^{(m,g)}e^{-\alpha d}}{1+
\tilde{r}^{(v,m)}{\vphantom{\tilde{r}^{(v)}}}^2
\tilde{r}^{(m,g)}{\vphantom{\tilde{r}^{(v)}}}^2
e^{-2\alpha d}{\vphantom{\tilde{r}^{(v)}}}-
2\tilde{r}^{(v,m)}\tilde{r}^{(m,g)}e^{-\alpha d}\cos\psi},
\label{eq14}
\end{equation}
\noindent
where the respective coefficients are given by
\begin{equation}
\tilde{t}^{(v,m)}=\frac{4n^{(m)}}{(1+n^{(m)})^2},
\qquad
\tilde{t}^{(m,g)}=\frac{4n^{(m)}n^{(g)}}{(n^{(m)}+n^{(g)})^2}.
\label{eq15}
\end{equation}

As  a result, for the absorptance of light by an atomically thin
metallic film of thickness $d$ we obtain
\begin{equation}
{\cal A}=1-{\cal R-T}.
\label{eq16}
\end{equation}

According to Sec.~III, the most often used metal in Casimir physics
is Au \cite{5,13}. However, rather low reflectivity of Au in the visible
light makes it unsuitable for using in Fabri-P\'{e}rot filter.
Thus, even at the boundary plane between an Au semispace and vacuum
${\cal R}_{ss}^{(\rm Au)}=0.71$. For the atomically thin films used below
($d$ varies from 0.94 to 1.41\,nm) the reflectance ${\cal R}^{(\rm Au)}$
computed using Eq.~(\ref{eq12}) varies from approximately 0.36 to 0.32,
i.e., is even much lower.

Because of this, we choose Ag as the metal of the mirrors in the proposed
setup. In this case Eq.~(\ref{eq12}) leads to
${\cal R}^{(\rm Ag)}=0.938$ for the film of $d=0.94\,$nm thickness and
to ${\cal R}^{(\rm Ag)}=0.930$ for the film with $d=1.41\,$nm.
The respective values of the film transmittance
${\cal T}^{(\rm Ag)}$ computed using Eq.~(\ref{eq14})
are 0.044 and 0.043. In this case the absorptance of light by one mirror
${\cal A}^{(\rm Ag)}$
computed by Eq.~(\ref{eq16}) is equal to 0.018 and 0.027, respectively.

Assuming that for both mirrors the values of  ${\cal R}^{(\rm Ag)}$ and
${\cal T}^{(\rm Ag)}$ are equal and that the power of light entering the
resonator is amplified by the factor of $q$ one finds for the transmission
coefficient \cite{65}
\begin{equation}
\tau=\left(1-
\frac{{\cal A}^{(\rm Ag)}}{1-{\cal R}^{(\rm Ag)}}\right)^2,
\label{eq17}
\end{equation}
\noindent
where $q=1/(1-{\cal R}^{(\rm Ag)})^2$ represents the quality factor in
our formalism.
The coefficient $\tau$ allows
calculation of the power of light transmitted through the resonator
$N_{\rm tr}= \tau N_{\rm in}$. For the two microresonators with mirror
thicknesses equal to 0.94 and 1.41\,nm we obtain from Eq.~(\ref{eq17})
$\tau=0.50$ and 0.38, respectively, i.e.,  $N_{\rm tr}=3.5$ and 2.66\,mW.
This means that for the two microresonators under consideration the
light power passed through the left mirror (it is equal to
$N_{\rm in}{\cal T}^{(\rm Ag)}=0.31$ and 0.30\,mW)
is amplified by the factors of 260 and 206, respectively.

Now we are in a position to determine the thickness of resonator mirrors
furnishing a balance between the Casimir and light pressures.
The pressure of light amplified in the resonator is given by \cite{65}
\begin{equation}
P_{\,\rm light}=\frac{1}{c}(1+{\cal R}^{(\rm Ag)})I_{\rm res},
\label{eq18}
\end{equation}
\noindent
where the amplified intensity is $I_{\rm res}=qI_{\rm in}{\cal T}^{(\rm Ag)}$.
Then the condition that the Casimir force acting on a SiO${}_2$ wall
is compensated by the force due to the light pressure takes the form
\begin{equation}
\int_S P_{\,\rm light}dS\approx
\left|P\left(\frac{\lambda}{2}+\Delta\lambda\right)\right|D_1^2,
\label{eq19}
\end{equation}
\noindent
where the integration is performed over the area of the light beam.

Substituting Eq.~(\ref{eq18}) in Eq.~(\ref{eq19}), one obtains
\begin{equation}
\frac{1}{c}(1+{\cal R}^{(\rm Ag)})N_{\rm res}
\approx
\left|P\left(\frac{\lambda}{2}+\Delta\lambda\right)\right|D_1^2,
\label{eq20}
\end{equation}
\noindent
where the power of light, amplified in the resonator, is
$N_{\rm res}=qN_{\rm in}{\cal T}^{(\rm Ag)}$.
We remind that $\Delta\lambda$ was chosen sufficiently large so that
for separation between the mirrors equal to $\lambda/2+\Delta\lambda$
the resonance condition breaks down. In our case
$\Delta\lambda=\lambda/q$.

By analyzing Eq.~(\ref{eq20}) with account of computational results for
the Casimir pressure, we find that this equality holds for $d=1.175\,$nm
(which corresponds to the metallic mirror containing five atomic layers).
In so doing, ${\cal R}^{(\rm Ag)}=0.934$, $q=230$, and $\Delta\lambda=2.3\,$nm.
This leads to the Casimir pressure in the vertical position of the wall
equal to $P=-180.4\,$mPa, i.e., to the attractive Casimir force
$F=PD_1^2=-0.45\,$nN and to the repulsive force of the same magnitude due
to the light pressure.

In the end of this section we briefly discuss
the role of various background effects, such
as electric forces due to a residual potential difference,
radiation friction, and bolometric forces. For metallic mirrors
used in the optical chopper the role of residual electric force
can be made negligibly small by the Ar-ion cleaning procedure
developed recently in application to the Casimir-based microdevices
in Ref.~\cite{68}. For the values of the light pressure considered
above the role of radiation friction and other optomechanical effects
remains negligibly small~\cite{51aa,51bb}. The bolometric force arises
due to light absorption. In optomechanical systems bolometric forces
may lead to a deflection of a specially optimized microlever having
the spring constant of approximately 0.01 N/m \cite{69}. Simple
calculation shows that to reach a deflection of the top of our wall
for 2.3 nm under the influence of the Casimir force calculated above,
the spring constant of a wall, supporting the right mirror, should
be equal to at least 0.1~N/m, i.e., by an order of magnitude larger.
The bolometric forces may make only a negligibly small impact on an
Al-coated wall with a relatively high resistance to tilting.
This makes inessential their possible role in the proposed device.

Creation of the Fabri-P\'{e}rot microfilter with the above parameters makes
it possible obtaining discrete pulses of the transmitted light   from a continuous
wave of incident light. For the found value of $d=1.175\,$nm, providing the
desired equality between the magnitudes of the Casimir force and the
light-pressure force, one obtains the transmittance
${\cal T}^{(\rm Ag)}=0.0436$, the light power entering the resonator
$N_{\rm in}{\cal T}^{(\rm Ag)}=0.305\,$mW, and the power of the amplified
light  $N_{\rm res}\approx 70\,$mW. The power of transmitted pulses
leaving the microfilter can be determined in two ways as
$N_{\rm tr}={\cal T}^{(\rm Ag)}N_{\rm res}$ or as $\tau N_{\rm in}$,
where, in accordance with Eq.~(\ref{eq17}), $\tau=0.44$, leading to the
common result $N_{\rm tr}\approx 3\,$mW.

\section{Conclusions and discussion}

In the foregoing, we have proposed the possibility to create
the novel microdevice driven by the vacuum
fluctuations of the electromagnetic field. This device includes
the Fabry-P\'{e}rot microfilter with two parallel atomically
thin metallic mirrors separated only slightly in excess of the
half wavelength of an incident light. One of these mirrors
should be deposited on the sufficiently thin wall subjected to
bending under the competing impact of the attractive Casimir
force and (in the presence of a continuous wave produced by the
source laser) of the repulsive force due to the light pressure.
As a result, the resonance condition alternatively obeys and
breaks down, and the filter resonator periodically produces the
pulses of transmitted light. Thus, the proposed microdevice can
work as an optical chopper do not using any kind of rotating
wheels usually employed in mechanically based choppers.

To demonstrate the feasibility of the proposed microdevice, we
have developed theoretical description of both the Casimir force
acting between atomically thin metallic mirrors deposited on
dielectric substrates and of the reflectivity properties in the
Fabry-P\'{e}rot microfilter formed by these mirrors. It should
be noted that calculation of both the Casimir force and  the
reflectance and transmittance of the
boundary surfaces in a microfilter
in the case of atomically thin mirrors is not trivial. For the
Casimir force, where several first Matsubara frequencies
contribute essentially at the separation considered, it is
necessary to take into account an anisotropy in the dielectric
properties of metallic mirrors. When calculating the reflectivity
properties of an atomically thin metallic films to visible light,
the use of the standard, Fresnel, reflection coefficients leads
to contradictions with the measurement data, and in this case
several phenomenological approaches have been developed
in the literature. We
have used the version of the Lifshitz theory adapted for
anisotropic materials to calculate the Casimir force and one
of such phenomenological approaches to calculate the
reflectivity properties in the Fabry-P\'{e}rot microfilter.
The specific values of all parameters of the proposed
microdevice have been determined which ensure its workability.

In the experimental realization of the proposed setup it is
desirable to perform control measurements of the reflectance and
transmittance of a unit wall made of quartz glass with
deposited atomically thin metallic films of various thickness,
as well as of its stiffness. This will help to confirm the
physical nature of all acting forces and the suitability of the
phenomenological approach used to calculate the reflectivity
properties.

To conclude, the proposed optical chopper driven by the Casimir
force may find prospective applications in the emerging area of
nanotechnological devices exploiting quantum fluctuations of the
electromagnetic field for their functionality.

\section*{Acknowledgments}
The work of V.M.M. was partially supported by the Russian
Government
Program of Competitive Growth of Kazan Federal University.

\newpage
\end{document}